# CAN A CYLINDRICAL GEOMETRY DESCRIBE DIFFUSION IN A NANOMETRIC POROUS MEDIA?

*P. C. T. D'Ajello*[*], *M. L. Sartorelli and L. Lauck*

*Universidade Federal de Santa Catarina*

*Departamento de Física/CFM*

*Brazil*

*P. O. Box 476 - CEP 88040-900*

*Fax 55 (48) 3721 9946*

*e-mail:* pcesar@fisica.ufsc.br

**ABSTRACT**

In a recent paper [1] we developed a theoretical model to describe current transients arising during electrochemical deposition experiments performed at the bottom of sub-micrometric cylindrical vessels with permeable walls. In the present work we extended the model for describing the current transients observed during electrodeposition through porous networks produced by colloidal crystals. Instead of considering a cylindrically shaped membrane with a constant cross sectional radius, the membrane will have a corrugated surface, with a radius that changes periodically with z, the vertical axis of the cylindrically corrugated vessel. According to the model, the porous network is formed by the replication of those units, put side by side in close contact, and impregnated by an electrolytic solution. Through the lateral surface of those cylinders we allow for a selective flux of species. The inward or outward flux obeys a complex dynamics regulated by the competition between the diffusion kinetics and the chemical kinetics that answer for the reduction of species at a reactive surface located at the bottom of the cylindrical cavities. The analytical expression for the current transient is complemented by a random prescription for the influx or outflux of matter through the lateral surface plus a modulation in its intensity that follows the surface corrugation. The theoretical data are compared with the current transients obtained in nanosphere lithography experiments.

## *1. INTRODUCTION*

   *This article presents a theoretical method for describing electrochemical deposition dynamics in nanometric porous media. The nanometric dimensions of the pores will set the special conditions of the system: the effect of capillarity is strong enough to prevent the action of gravitational forces and there is no movement of the fluid as a whole, i.e., there is no drift velocity. The fluid stays in quiescent conditions and the generalized chemical forces are the sole responsible for driving species throughout the liquid medium. This situation is typical in electrochemical deposition through colloidal masks made of nanospheres, to mention an experimental situation that motivates our theoretical work.*

   *In the experimental array whereupon we focus our attention, the porous system is composed of submicrometer spheres ordered in a fcc structure, on top of a flat electroactive substrate [2,3]. The electrolytic medium fills the interstices among the spheres. The geometry dictated by the nanosphere's close packed array, plus a heterogeneous reaction, produce a material deposit that grows on top of the electroactive surface and fills the interstices among the spheres. In Figure 1 we show the experimental current transients obtained during cobalt electrodeposition through monolayered colloidal polystyrene masks [3]. The diameters of the polystyrene spheres are indicated in the figure and will correspond to 2R in the theoretical description. The masks made with 165 and 600 nm spheres where nearly perfect monolayers, whereas the mask made with 496 nm spheres had regions partially covered with a second layer of spheres. In all curves the point of minimum current is supposed to correspond to a film thickness that equals the sphere's radius, i.e., when the film/liquid interface is at the hemispherical equatorial plane. For the 165 and 600 nm masks it is seen that after complete*

*coverage of the sphere's monolayer the current yield again coincides with the current obtained in a flat substrate of same area. For the 496 nm mask the current limiting value is slightly lower than that, what is a direct consequence of inhomogeneities in the mask thickness.*

*From a theoretical point of view this class of system (with one, two, three or n layers), is identified as a prototypical porous medium, where particles, diluted in a liquid quiescent medium, diffuse toward a reactive surface.*

*The model tries to capture the essential features of the porous system, reducing it to a problem with cylindrical symmetry, amenable to analytical calculations. The following line of reasoning is considered. A colloidal crystal that is self-ordered on top of a flat substrate has its (111) axis directed perpendicular to the substrate, as shown in the top view of Fig. 2a. Along the (111) direction there is no straight path to the bottom and the ions are forced to deflect laterally in order to reach down the next layer of spheres. In Fig. 2a the curved white arrows single out one possible diffusion path for an ion that strikes the location marked with a star. The same diffusion path is shown by solid black arrows in the schematic vertical cross section of the porous structure (Fig. 2b). This particular path will be modeled as a staircased cylindrical vessel. At every inflection point along the twisted cylinder the diffusing particle may either remain in the same vessel or diffuse out into one of two other neighbouring units, what is indicated by the dotted white arrows in Fig. 2b. Conversely, at each inflection point a cylinder may receive particles coming from two other vessels. The porous structure, as a whole, can be seen as a periodic replication of a twisted cylinder, conveniently placed along the x and y direction, as pictured in Fig. 2c. As we see, the essence of the lateral diffusion of ions is to allow for exchange of particles among neighbouring vessels and this will be described by a periodic function g(z) that modulates, along the length of the cylinder, the*

*intensity of influx/outflux of particles across the walls, plus a random variable that defines the flux direction at each point of exchange. With these provisions two essential features of the problem are taken into account, namely, the lateral diffusion of ions and its periodic and randomic nature. Therefore, the twisted shape of the cylinders may now be neglected and the problem reduces to an array of porous and straight cylinders directed along the z axis, with a lateral periodicity that mirrors the hexagonal lattice seen at the surface of the porous structure.*

*The next point of analysis concerns the description of the internal wall of the cylindrical vessel. In real systems the cross-sectional area of a diffusion path has a complex shape. Along the z direction its open portions, where exchange of particles is possible, are periodically intercalated with closed portions of minimal cross-sectional area that are delimited by the closest approximation among three spheres. Moreover, the lateral exchange of particles has an angular dependence with a trigonal symmetry. The model, however, will ignore those details and focus solely on the periodicity of the cross sectional area exhibited along the z direction, assuming for it a cylindrical symmetry, with no angular dependence in the horizontal plane. The internal cross sectional radius of the vessel will vary, in a continuous and periodic fashion, from a minimum value $R_{min}$ up to R, which is the external radius of the cylinder, as shown in Fig. 2d. The need of assuming a corrugated internal surface for the cylinders will become more evident later on.*

*In what follows, we will solve the diffusion problem for one single cylindrical vessel. We begin with an analytical expression recently derived [1] to describe the current transients recorded during the heterogeneous nucleation and deposition of ions at the bottom of a right circular cavity.*

The system is prepared satisfying a low concentration condition for species in a homogeneous and stationary fluid. We also assume that when a potential is turned on, species start to react at the electrode surface placed at the bottom of the cylinder. As a consequence, a concentration gradient sets up giving rise to a generalized force that guides the ionic species toward the reactive surface. The species flow but the solvent, in which they are diluted, remains under quiescent conditions. Thus, in our model the diffusion coefficient remains constant because the temperature, the magnitude of the particles and the fluid viscosity are assumed to remain constant along the deposition experiment.

The paper is organized as follows: in Section 2 we introduce the model and the solution obtained for a cylindrical vessel with straight walls. In Section 3 we examine the current transient profiles of a cylindrical cavity with random influx/outflux of species, directed normal to the lateral surface of the cylinder, with a modulated intensity along the z axis. We also discuss the need of introducing corrugated walls in the cylinder in order to reproduce the experimental data. In Section 4 the proposed model is applied to the experimental curves shown in Fig. 1. In Section 5 we present some comments and conclusions.

## 2. The model

We look for an expression for the current produced by the reduction and deposition of ions at the electrode surface. The electrode surface lies at the bottom of a cylindrical cavity. The cylindrical cavity is filled by an electrolytic solution that contains the ionic species, homogeneously diluted at the initial time. The ionic concentration inside the finite cylindrical region obeys the balance equation

$$\frac{\partial C(r,\theta,z,t)}{\partial t} = D\nabla^2 C(r,\theta,z,t) \quad . \tag{1}$$

In this equation, $C(r,\theta,z,t)$ is the local ionic concentration, $t$ is the time and $D$ is the diffusion coefficient of the solvated ions. Because the electrode surface is homogeneously reactive only at the bottom of the right circular cavity, we assume rotational symmetry around the vertical axis of the cylinder and Eq. (1) can be simplified to yield the balance equation in cylindrical coordinates

$$\frac{\partial C}{\partial t} = D\left[\frac{\partial^2 C}{\partial r^2} + \frac{1}{r}\frac{\partial C}{\partial r} + \frac{\partial^2 C}{\partial z^2}\right], \tag{2}$$

where $z$ is the position in the direction normal to the reactive surface, located at $z = 0$.

To solve Eq. (2), we adopt the following initial and boundary conditions:

$$C(r,z,0) = c_b \qquad R \geq r \geq 0, h \geq z \geq 0, \tag{3a}$$

$$C(r,0,t) = c_b e^{-kt} \qquad R \geq r \geq 0, \forall t \geq 0, \tag{3b}$$

$$C(r,h,t) = c_b \qquad R \geq r \geq 0, \forall t \geq 0, \tag{3c}$$

$$\left.\frac{\partial C(r,z,t)}{\partial r}\right|_{r=0} = 0 \qquad R \geq r \geq 0, \ h \geq z \geq 0, \forall t \geq 0, \tag{3d}$$

$$\left.\frac{\partial C(r,z,t)}{\partial r}\right|_{r=R} = -\alpha c_b\left(1 - e^{-\nu t}\right)g(z) \qquad R \geq r \geq 0, h \geq z \geq 0, t \geq 0. \tag{3e}$$

*The initial condition (3a) asserts that, before the electric potential is turned on, the ions are uniformly distributed in the region inside the cylinder, and the ionic concentration equals the bulk concentration $c_b$. The time-dependent boundary condition (3b) determines the temporal evolution of the ionic concentration at the electrode surface. This condition is regulated by the magnitude of $k$, which describes the reactive properties of the electrode. The parameter $k$ describes the rate at which the surface transfers electrons to reduce the ions. Therefore, $k$ is a factor that defines the reaction kinetics. Despite its simplicity, boundary condition (3b) plays a central role in our description because it defines the concentration drop at the electrode where the ions are reduced and withdrawn from the liquid medium. If $k = 0$ there is no reaction and there are no concentration changes at the surface ($C = c_b$ at any time). For $k \neq 0$, the concentration at the electrode surface evolves to a stationary value (when $t \to \infty$). A concentration gradient is produced and ions migrate from the bulk of the solution toward the surface. The boundary condition (3c) guarantees that we are working with a cylinder whose length is at least equal to the stationary diffusion layer that defines a distance from the electrode surface, beyond which the ion concentration is assumed to be constant and equal to the bulk concentration $c_b$. Boundary condition (3d) is a consequence of the rotational symmetry of our system. Finally the time dependent boundary condition (3e) determines the evolution of the concentration of ions that flow through the lateral surface of the cavity ($R$ is the outer cylinder radius), in its normal direction. This boundary condition also depends on $z$, i.e., the flux of matter through the cavity's lateral surface can change along its length. This boundary condition contains the parameter $\alpha$, which quantifies the magnitude of matter flux that crosses the cylindrical lateral surface. Its signal determines the direction of matter flow, inward for positive $\alpha$ or outward if $\alpha$ is negative. The time*

*dependent expression contained inside the brackets is a mathematical requirement to guarantee consistency of the boundary and initial conditions. Thus at $t = 0$ there is no flow and the system is characterized by a constant and homogeneous distribution of matter ($C(r,z,t) = c_b$). When the potential is switched on, the symmetry is broken off. Ions begin to react at the cylinder bottom and a gradient arises to guide the species toward the electrode surface. In addition, the flow across the lateral area of the cylinder obeys a transient rule, quantified by the magnitude of the rate constant $\nu$ that appears in the exponential argument of Eq. (3e). A physical reasoning to justify the temporal dependent term in boundary condition (3e) is that $\nu$ quantifies the time interval elapsed until an external pumping mechanism (osmotic pressure for example) attains its maximum value according to the rule prescribed by $g(z)$ on the lateral surface of the cavity. Regardless of this interpretation, we wish to emphasize the relevance of boundary condition (3e) for our model. It is the condition that sustains the plasticity of the model, namely, its capacity to reproduce different situations, according to the signal of $\alpha$ and the form of function $g(z)$. Thus a careful choice of parameter $\alpha$ and function $g(z)$ yields a good description of the diffusion phenomenon and heterogeneous reaction in a porous medium.*

*To support its general character we do not specify the form of the function $g(z)$ from the beginning.*

*To solve the problem we first perform a Laplace transformation on the differential equation and its boundary and initial conditions, then we develop a sine Fourier transformation over $z$, with $0 \le z \le h$. A detailed development of these procedures is given in reference [1], where we show that after the inverse transformations we arrive at the solution,*

$$C(r,z,t) = c_b - 2c_b \sum_{n=1}^{\infty} \left( \frac{1 - e^{-\omega_n^2 t}}{n\pi} \right) . sin\left( \frac{n\pi}{h} z \right) + 2c_b \sum_{n=1}^{\infty} \frac{\omega_n^2}{n\pi} sin\left( \frac{n\pi}{h} z \right) \left[ \frac{e^{-\omega_n^2 t} - e^{-kt}}{(k - \omega_n^2)} \right] + \frac{4\alpha R c_b}{h\gamma^2} .$$

$$\sum_{n=1}^{\infty} \widetilde{g}_n . sin\left( \frac{n\pi}{h} z \right) \left\{ \begin{array}{c} \dfrac{J_0\left( i \dfrac{r}{R} \gamma \omega_n \right)}{\omega_n^2 \left[ J_0(i\gamma\omega_n) + J_2(i\gamma\omega_n) \right]} - \dfrac{e^{-\nu t} J_0\left( i\gamma \dfrac{r}{R} \sqrt{\omega_n^2 - \nu} \right)}{(\omega_n^2 - \nu) \left[ J_0\left( i\gamma \sqrt{\omega_n^2 - \nu} \right) + J_2\left( i\gamma \sqrt{\omega_n^2 - \nu} \right) \right]} + \\ \\ \dfrac{\nu . e^{-\omega_n^2 t}}{\omega_n^2 (\omega_n^2 - \nu)} - 4 \sum_{\lambda_m} \dfrac{\nu e^{-x_{n,m}t} J_0\left( \dfrac{r}{R} \lambda_m \right)}{\left( \omega_n^2 + \dfrac{\lambda_m^2}{\gamma^2} \right) \left( \omega_n^2 - \nu + \dfrac{\lambda_m^2}{\gamma^2} \right) \lambda_m J_3(\lambda_m)} \end{array} \right\} ,$$

*(4)*

where, $J_0$ and $J_1$ are the zeroth and first order Bessel functions and $\gamma$, $\beta$, $\omega_n^2$ are

$$\gamma = \frac{R}{\sqrt{D}},$$  *(5)*

$$\beta = \frac{r}{R},$$  *(6)*

and

$$\omega_n^2 = \frac{n^2 \pi^2 D}{h^2}.$$  *(7)*

In Eq. (4), $\lambda_m$ is the $m^{th}$ root of the first order Bessel function, $J_1(\lambda_m) = 0$ and

$$\widetilde{g}_n = \int_0^h sin\left( \frac{n\pi}{h} z \right) g(z) \ dz.$$  *(8)*

Because we search for an expression that allows for a comparison between theoretical and experimental data we calculate the current density,

$$J(r,z,t) = -D\left[\frac{\partial C}{\partial r}\hat{e}_r + \frac{1}{r}\frac{\partial C}{\partial \theta}\hat{e}_\theta + \frac{\partial C}{\partial z}\hat{e}_z\right].$$

*In this equation only the last term is relevant for us, because it describes the flux at the reactive surface at the bottom of the cylindrical cavity. The second term is zero, because of the rotational symmetry. The first term, even being non zero, does not contribute to the electric current density because there is no charge flux (to the electrode) in the radial direction, just flow of matter. Thus the current density reads:*

$$J_z(r,z,t) = -\bar{z}FD\frac{\partial C(r,z,t)}{\partial z}, \tag{9}$$

*where $\bar{z}$ is the charge number in a particular reduction process, $F$ is the Faraday constant and $C(r,z,t)$ is given by Eq. (4). We observe that Eq. (9) is the longitudinal component of the electric current transient.*

*An expression for $J_z(r,z,t)$ is easily obtained deriving Eq. (4) in the variable $z$, although the resulting equation is very cumbersome because the rates of Bessel functions appearing in the last term of Eq. (4) still remain. To avoid this problem we perform the integration of the current density on the very base of the cylinder. This procedure gives us the total current that flows through the electrode surface, with the advantage of reducing the number of variables and simplifying the current equation. Additionally, it allows for comparison between theoretical and experimental data. In this case,*

$$I(z,t) = \int_0^R J_z(r,z,t)\,2\pi r\,dr, \tag{10}$$

*gives the total current that flows into the reactive base of the cylindrical cavity,*

$$I(z,t) = \frac{2\bar{z}DFc_b \, \pi R^2}{h} \left[ -\sum_{n=1}^{n_{max}} \left( \cos\frac{n\pi}{h}z \right) \left( 1 - e^{-\omega_n^2 t} \right) - \sum_{n=1}^{\infty} \frac{\omega_n^2}{(k-\omega_n^2)} \left( \cos\frac{n\pi}{h}z \right) \left( e^{-\omega_n^2 t} - e^{-kt} \right) \right]$$

$$- \frac{2\bar{z}DFc_b \, \pi R^2}{h} \left[ 2\frac{\alpha h}{R} \sum_{n=1}^{n_{max}} \frac{\tilde{g}_n}{n\pi} \left( \frac{\omega_n^2}{\omega_n^2 - \nu} \right) \left( \cos\frac{n\pi}{h}z \right) \left\{ 1 - e^{-\nu t} - \frac{\nu}{\omega_n^2} \left( 1 - e^{-\omega_n^2 t} \right) \right\} \right].$$

*(11)*

We observe that the sum index runs in a finite interval, which is explained in reference [1]. Equation (11) can also be expressed as:

$$I(z,t) = \frac{2\bar{z}DFc_b \pi R^2}{h} f(t),$$ (12)

Where $f(t)$ is the time dependent expression in brackets.

## 3. Results

In this section we examine some results produced by our theoretical model. Our goal is to describe the current transient of species that react at the bottom electrode after diffusion through a porous medium in a quiescent liquid medium.

To emphasize the relationship between the current profiles and the geometric shape of the porous system, we first consider a cylindrical cavity with flat walls and an inward/outward flux of species through them. This situation fits perfectly the model we *have developed, with sources or sinks at the lateral cylindrical surface obeying rotational symmetry, as in reference [1], but differing on the intensity and the direction of the ionic flux.*

*In a previous work [1] we considered three cases involving a cylindrically shaped cavity, i.e.: (i) $g(z) = 0$, which acts as a cavity surrounded by an impermeable membrane, (ii) $g(z) = 1$ and $\alpha = c > 0$ that corresponds to a semi-permeable membrane with a constant influx of species and, finally, (iii) $g(z) = 1$, $\alpha = c < 0$, which describes a cylindrical cavity with semi-permeable membrane and a constant outward flow of species in the radial direction. Now, in order to describe the porous medium, with radial flux of matter (species), a non-constant $g(z)$ function is assumed:*

$$g(z) = \beta \ \cos^2\left(\frac{\pi}{2R} z\right),$$ *(13)*

*that defines a periodic behavior for the flux of species that enter or leave the cylindrical cavity according to its position in space. There is still a rotational symmetry and also a periodicity, once $g(z) = 0$ for $z = qR$ with $q = 1,3,5,....$ and $g(z) = 1$ when $q = 0,2,4,....$.*

*We still have a rigorously defined cylindrical cavity, but the flux of species obeys a spatial distribution that identifies the regions where species have easy access to the cylindrical cavity (the pore). Certainly the direction of the lateral flow, defined by the signal of $\alpha$, has a complex dynamic behavior. Thus we adopt the simplest proposition. We assume that the signal and the magnitude of $\alpha$ are random variables with a null ensemble average, assigning to $\alpha$ a random walk behavior.*

*To introduce the random behavior for α we use a Monte Carlo algorithm, applied in Eq. (11), based on the plain random walk in the space of this parameter, in order to*

account for the fact that at each layer the particle may stay in the same vessel or diffuse out into one of two neighboring vessels. When the particle takes one particular path, it is in fact withdrawn from the other two, a situation that undergoes inversion in a future time interval. This behavior on intensity and direction of the flow through the lateral surface of the cavity is represented by the product of functions $\alpha g(z)$, that regulates the representative horizontal bars appearing on Fig.2d. The stochastic character assumed by $\alpha g(z)$ is made effective operating on $\alpha$, which is done according to the following prescription: we begin with the initial value $\alpha_0 = 0$, defining a constant increment $\pm \Delta \alpha$, with the signal taken at random. If at a certain instant $t_1$ we define the parameter as $\alpha_1 = \alpha_0 \pm \Delta \alpha$ after a random choice of sign, this value is maintained for a time interval $\Delta \tau$, after which a new $\alpha_2 = \alpha_1 \pm \Delta \alpha$ is chosen at $t_2 = t_1 + \Delta \tau$, that will last for another time interval $\Delta \tau$. The procedure is repeated until the current transients are concluded.

There is a final aspect to consider before the application of the method. As the heterogeneous reaction that occurs at the electrode results in a electrodeposit, there will be a progressive increase in the deposited film thickness and consequently a progressive shift on the g(z) function that mimics the lateral flux. This effect changes the flux geometry near the topmost deposited layer. To account for that, the reactive surface is kept fixed at $z = 0$, whereas $z$ is shifted by a discrete and constant quantity, so that we have $g(z')$ with $z' = z + \Delta z$ where $\Delta z$ is the increment on the deposit thickness after a given time interval $\Delta t$. This trick allows us to compute Eq. (11) at $z = 0$ taking into account the correction in $\tilde{g}_n$. The increase $\Delta z$ can be computed per unit area by the expression:

$$\Delta z = \frac{2Dc_b}{h} \frac{M}{\rho} \int_t^{t+\Delta t} f(t)\, dt \times K \qquad (14)$$

*that uses the density ρ and the molar mass M of the ion that is being deposited. The constant factor K accounts for a correction term that might be necessary to fit the experimental data.*

*In Fig. 3 we use Eq. (11) to generate current transient profiles, using the standard values and $k = 0.89 \ s^{-1}$. We also assume that the cylinder radius is $R = 300 \ nm$ and the magnitude of the depletion layer $h$, i.e., the perpendicular distance from the reactive surface, beyond which the concentration of species remains constant and equal to the initial concentration $c_b$ is $h = 3x10^{-3} cm$. For all calculations we use a limited sum index given by $n_{\max} = 4000$. Each curvein Fig. 3 corresponds to a different seed for the random number generator, so that there is a different sequence of random numbers that define the signal and magnitude of $\alpha$ in Eq. (11). The instabilities observed in the current transients reflect the randomic nature of α. The average of the four curves yields a flat curve, as can be seen in the inset of Fig. 3. Through Fig. 3 it is easy to verify that the current has an undefined pattern, that do not resemble the ones shown in Fig. 1, and if we add all of them we get the flat curve, characteristic of a compact film deposition. This results indicates that the product of functions $\alpha g(z)$ is not enough to represent the real porous system. In what follows we will add a periodic corrugation at the lateral walls of the cylinder in order to try to mimic the porous geometry in an effective way.*

*To implement this idea we consider that $\pi R^2$ the cross sectional area of the cavity, is now a periodic function of z, described by $\pi \tilde{R}^2$, as sketched in Fig. 2d.*

$$\pi \tilde{R}^2 = \pi \left( R - 0.25R \left[ 1 - \cos^2 \left( \frac{\pi}{2R} z \right) \right] \right)^2. \qquad (15)$$

*The function $g(z)$ has still the form given by Eq.(13) and $\alpha$ obeys the same prescription as before. Figure 4 presents the theoretical results for a $2R = 600\ nm$ unit cellular cavity, assuming that:*

$$g(z) = \beta \cos^2\left(\frac{\pi}{2R}z\right) \qquad 0 \le z \le 2R, \qquad \pi\widetilde{R}^2 \quad given \ by \ Eq.(15)$$
$$g(z) = 0 \qquad\qquad\qquad z > 2R,$$

*The curves described by symbols correspond to current transients obtained from Eq. (11), using the same values applied to generate the curves of Fig. 3. The continuous black curve is an average of the four curves. It is observed that the current now displays a broad minimum before reaching the stationary value. The results shown in Fig. 4 demonstrate that corrugations along the diffusing path are indeed necessary to describe the characteristic features observed in the experimental current transients.*

*In Figure 5 we show three simulated transients corresponding to unitary cells that differs only by the magnitude of R. The curves were obtained with the same seed used to generate the curve drawn with open circles in Fig. 4. Because the current profiles in Fig. 5 must correspond to the experimentally observed ones, shown in Fig. 1, where despite the volume of the pore, all of them fill the same volume with an equal reacting area, we must correct the number of cylindrical unities in each case. Thus the current profile obtained for $R = 248\ nm$ was multiplied by $1,463$ to yield the curve drawn with open circles in Figure 5. Likewise the current profile generated when $R = 82.5\ nm$ was multiplied by $13,223$ in order to give the curve sketched by a continuous line that represents the current in an electrolytic cell with the same reaction area like the others. A comparison between Figure 5 and Figure 1 shows that the model is able to describe the experiment reasonably well, but there is still room for improvement.*

*A first aspect to consider is the fact that the curves shown in Fig. 5 represent just one vessel, whereas the current transients obtained in the experiment represents the average behavior over $10^8 - 10^9$ vessels placed side by side. Furthermore, when dealing with systems composed of many layers of spheres, the correlation length among vessels should become an important issue. According to our description, the vessels are able to exchange particles with two other different vessels in each level, establishing a braided type of network. As time evolves, the correlation among vessels increases and we do not know how to treat this problem in a proper way.*

*Fortunately, there are two other important aspects that spare us from dwelling on the complexity sketched in the previous paragraph. For very large systems, like the one represented by the experiment, a statistical description is necessary. At the point of exchange with other vessels, the diffusion ion has a 33% chance of remaining in the same vessel against 66% chance of escaping to the neighboring ones. But each vessel has also a 66% chance of receiving particles from neighboring units. Therefore, on the average, the number of diffusing particles in each vessel remains constant, and in this sense, the parameter $\alpha$ can be taken as zero. Furthermore, the model itself has shown that, by assuming a corrugated wall, the role played by the product $\alpha g(z)$ becomes negligible, because the periodic change in the cross sectional area of the cylinder dominates the whole process.*

*The arguments introduced in the previous paragraph could give the idea that the existence of permeable walls and a randomic flux of particles among neighboring vessels turns out to be irrelevant. This is a precipitated conclusion given the actual stage of development of our model. It is enough to remember that even when $\alpha$ is a random*

*variable with an ensemble average equal to zero, it also depicts a square deviation that is, on the average, different from zero, and this will influence the final result when the statistical computation of all the units that form the system are adequately considered.*

*In spite of this warning, in the final section we will apply the model, assuming $\alpha = 0$ , for an examination of the experimental data shown in Fig. 1.*

### 4. Application of the model to a real porous system

*It is our purpose now to compare experimental curves with the proposed model in order to gain some insight on the physical mechanisms that underlie the process of ionic diffusion in nanometric porous media. It is clear that only a qualitative description will be possible, given the many simplifications assumed in the model.*

*The current transients of Fig. 1 were obtained from electrodeposition of cobalt through monolayered masks self-arranged on top of flat (100) n-Si. The exposed area was 0.496 cm$^2$ and the concentration of electroactive ions in the electrolyte was 0.305 mM. Further details are given in [3]. In order to approximate experiment and theory, the current transients will be normalized as follows: (i) the experimental current transients will be normalized by the area of the electrode; (ii) the theoretical curves, evaluated at z=0, will be normalized by $\pi R^2$; (iii) additionally, the theoretical expressions will also be normalized by $n_{max}$, which is an extrinsic multiplicative factor that arises in Eq. (11) from derivation of Eq. (4), and as such will be considered to have no physical meaning [1].*

As already explained before, we will consider that the vessels have impermeable walls ($\alpha = 0$). Therefore, the parameters $\nu$ and $g(z)$ do not have to be specified. Values for the reaction rate $k$ and for the ratio ($D/h$) will be extracted from the normalized current transient that was measured at a flat electrode (Fig. 1), what, according to the model is proportional to:

$$\lim_{t \to \infty} I(z=0) = 2\bar{z} \frac{D}{h} F c_b$$

The values that best describe the experimental curve are $k = 0.15 \ s^{-1}$ and ($D/h$) = $3.43x10^{-5}$ cm/s. A comparison between both curves is shown in Fig. 6a and one observes that the agreement is quite good. These values will be used in the description of the current transients obtained at porous electrodes. However, independent values for $D$ and $h$ are not available for the present system, therefore, we will assume a fixed value for $h = 3. \ 10^{-2}$ cm, based on typical values found in literature [4]. This assumption implies a value for the diffusion coefficient $D = 1.03 \ 10^{-6} \ cm^2 s^{-1}$, which is also a reasonable value, similar to the ones found in literature for aqueous electrolytes [2,5]. These values will be used in Eq. (14) to determine the increase, per unit area, in the height of the deposit after an interval $\Delta t$. It is also necessary to find a value for the numerical factor $K$ that best describes the current transients measured at the porous electrodes. Variations in $K$ affect solely the time scale, i.e., the instant at which the minimum in the current transient is observed. Surprisingly, a single numerical value, $K = 7.95x10^{-4}$, was able to describe all current transients, as can be seen in Fig. 6 (b-d). The qualitative agreement is very good, if one considers the fact that only one ad-hoc parameter was used to adjust all curves. There are, however, two important features observed in the experiments that are not being correctly described by the present model.

First, one observes in Fig. 1 that the absolute value of the current measured at its point of minimum increases with the radius of the spheres. This experimental evidence indicates that geometry is not the sole factor that regulates the current transient. There is a physical process, not accounted for by the model, that affects differently porous systems of different sizes. Second, the model is not able to predict correctly what happens when a second layer of spheres is added on top of the first one. As shown in Fig. 7, that depicts the current transients obtained for one and two layers of spheres, the first minimum in both curves occurs at the same time. However it was already demonstrated, for systems made with 600 nm spheres, that the presence of just one additional layer reduces the amount of deposited material to just 8 % of what is deposited, in the same period, through a single monolayer. This is a very strong reduction in the rate of deposition that is not explained neither in this model or elsewhere [2].

Finally, it is also important to comment on the value found for the correction factor $K$. According to Eq. 14 it can be seen as a factor that scales down the quantity (D/h), <u>inside</u> the porous medium. This is an indication that the model needs improvement, i.e. the assumption of a constant rate (D/h) throughout the process does not mirror reality. As described above, the ratio (D/h) was determined from the current density plateau measured at a flat electrode, where ionic diffusion proceeds without constraints. However, at the point of minimum current density, the reacting surface is immersed in the porous medium, where ionic diffusion is restricted by the proximity of walls. Reduced diffusivities in porous system are a well known experimental fact that is handled, theoretically, by the introduction of empirical parameters like tortuosity, constrictivity and porosity [2,5-10]. In a somewhat different approach, our efforts in the future will be directed toward modifying the boundary and initial conditions that set up the diffusion and reaction problem in a *corrugated cylindrical geometry, in order to correctly describe the wealthy of information*

*provided by the experimental model system.*

## *4. Comments and conclusions*

*In this paper we demonstrated that an ordered porous system can be represented by a set of cylindrical vessels with permeable walls. A set of boundary and initial conditions for diffusion and reaction in a cylindrical vessel results in an analytical expression for the current transient measured at the bottom of the cylinder that is completely defined by a group of parameters that control the intensity of ionic flux through the walls and the rate of ion consumption at the bottom of the cylinder. The model was used to describe experimental data recorded during electrodeposition on flat electrodes modified by colloidal crystals. It was shown that the introduction of a periodic corrugation in the cylindrical wall is essential to reproduce the current minima observed in experiments. It was also shown that for very large systems, the permeabilitiy of walls may be neglected. Results indicate, however, that despite the good qualitative agreement obtained with experimental data, the model needs to be improved in order to explain some features observed in real systems.*

*Acknowledgements*

*The authors would like to acknowledge to the Brazilian agency CNPq, for financial support, and in particular, the INCT of Organic Electronics.*

**Figure Captions**

**Figure 1:** Current transients obtained during cobalt electrodeposition through monolayered polystyrene colloidal masks, after [3]. The diameters of the polystyrene spheres are indicated in the figure. The solid line represents the current transient measured at a flat electrode.

**Figure 2:** Can a cylindrical geometry describe an ordered porous media? (a) Top view of a colloidal crystal formed by four layers of spheres self-ordered on a flat substrate in a fcc structure. The star at the center indicates one pore through which an impinging ion may enter the porous structure. The white arrows depict one possible diffusion path towards the flat substrate. (b) Schematic cross section of the colloidal crystal. Black arrows indicate the diffusion path that was singled out in (a). (c) That particular diffusion path is now modeled as a staircased cylinder. The whole porous structure can be seen as a periodic replication of aligned twisted vessels. At each inflection point the vessels may exchange particles with neighboring ones. (d) In the last simplification step the twisted vessels transform into straight cylinders with corrugated walls. The function $g(z)$ is sketched on the right, indicating the points of maximum and minimum flux among vessels.

**Figure 3:** Total current flowing through the reactive base of the cylinder. The curves were obtained from Eq. (11) assuming $D = 1x10^{-5} cm^2 s^{-1}$, $k = 0.89 \ s^{-1}$, $\nu = 0.1 \ s^{-1}$ and $\delta = 0.25$. $\alpha$ is a random parameter, as described in the text and there is no corrugation on the cylindrical surface; $g(z)$ is given by Eq. (13). Each current profile was generated by a different sequence of random numbers.

**Figure 4:** Current transients obtained from Eq. (11) using the same conditions and seeds used to generated the current profiles shown in Fig. 3, except by a change in the effective reaction area, following Eq. (15). $g(z)$ is given by the same prescription used to generate Fig. 3 but now in a corrugated cylindrical vessel.

***Figure 5:*** *Theoretical current transients obtained from Eq. (11) that describe diffusion in one corrugated vessel with different radii, as indicated in the figure. The current profiles where generated by one and the same seed. In order to compare systems with the same electroactive area the curve in open circles was multiplied by* 1,463 *, whereas the profile drawn by a continuous line was multiplied by* 13,223.

***Figure 6:*** *Experimental and theoretical current density profiles. The former ones refer to the same data presented in Fig. 1. All the theoretical curves were generated from Eq. (11) and Eq. (15), with one single fitting parameter K= 7.95 x 10⁻⁴ (defined in Eq. (14)). In this exercise, the cylinders were assumed to have impermeable walls ($\alpha = 0$). Other parameters were:* $D = 1.03 \times 10^{-6} cm^2 / s$; $h = 3 \times 10^{-2} cm$ *and* $k = 0.15 \ s^{-1}$. *The different radii are indicated in the graphics.*

***Figure 7:*** *Theoretical current density profiles measured at the bottom of an impermeable cylinder with R = 300 nm. The solid curve represents a cylinder with one corrugation. Open symbols represent a cylinder with two corrugations.*

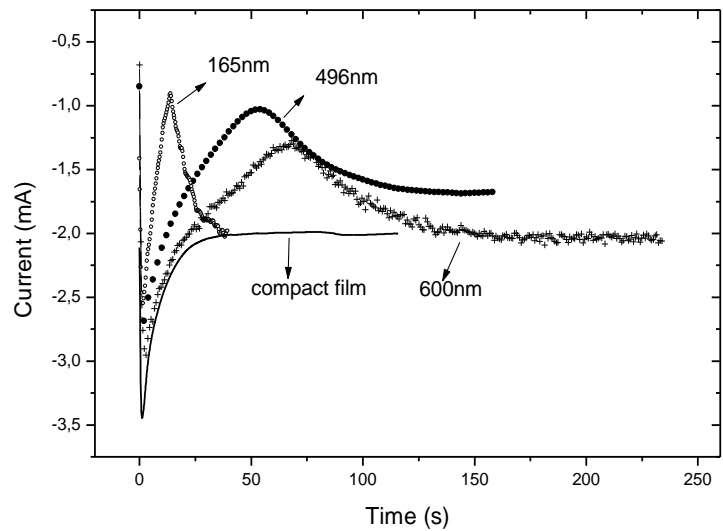

Figure: 1

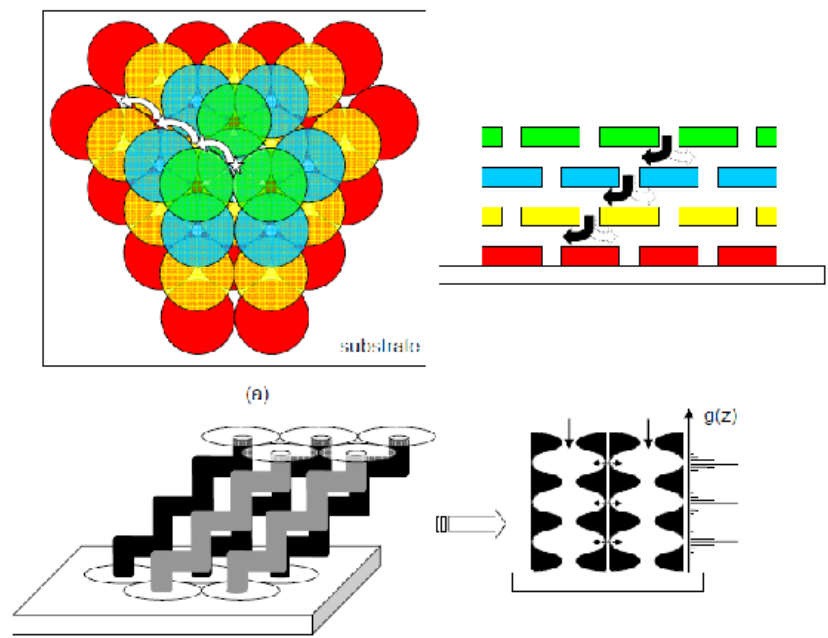

(a)

Figure: 2

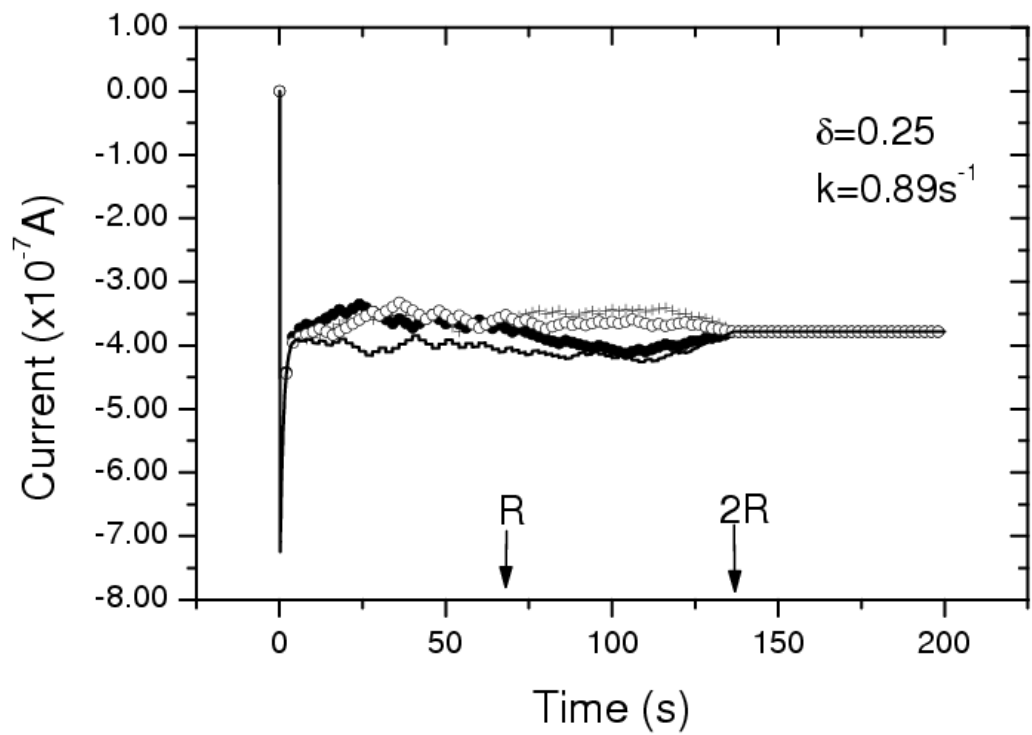

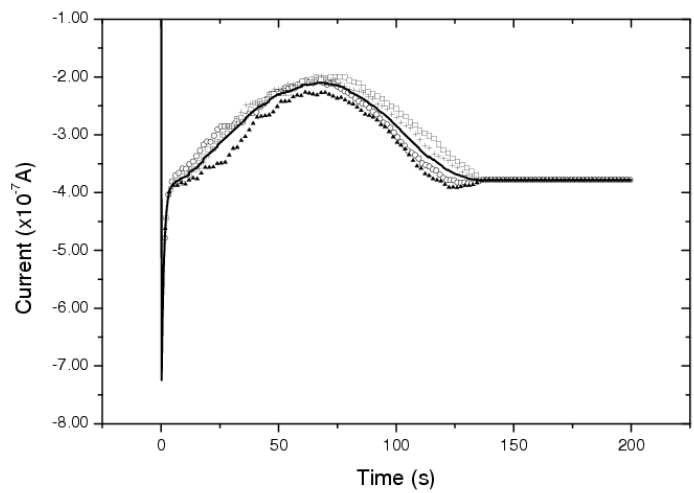

Figure: 4

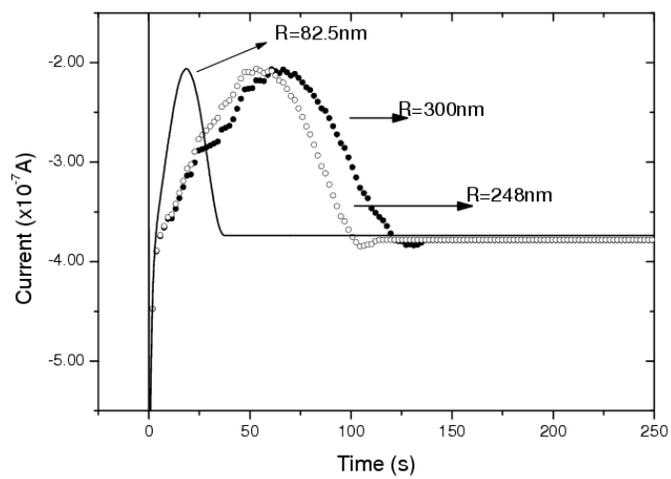

Figure:5

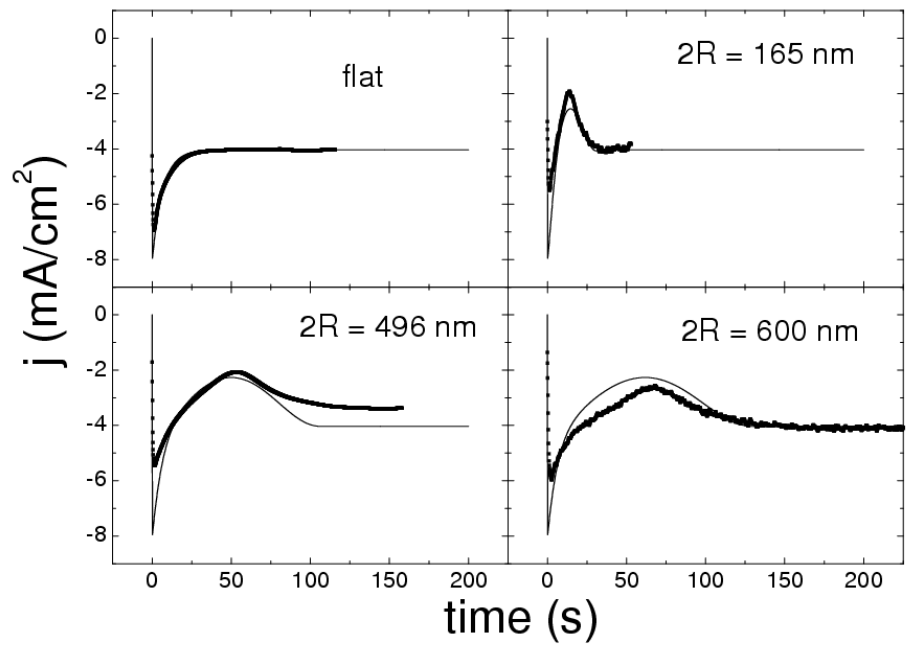

Figure:6

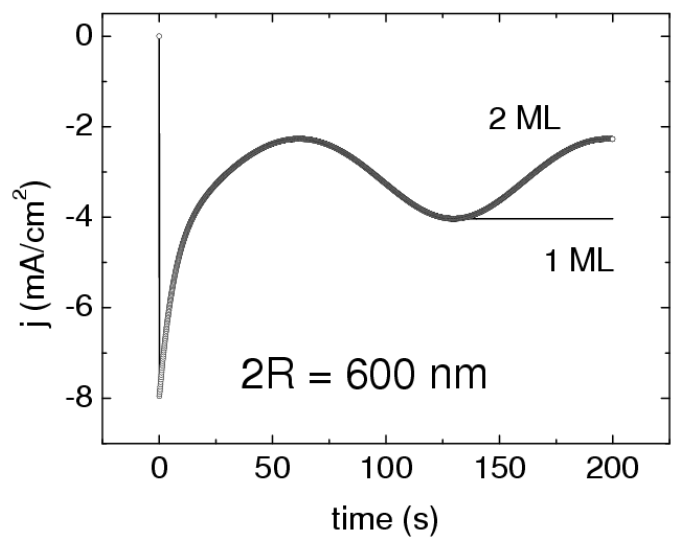

Figure:7